\documentclass[%
 reprint,
 amsmath,amssymb,
 aip,
]{revtex4-2}

\usepackage{graphicx}
\usepackage{dcolumn}
\usepackage{hyperref}
\usepackage{xr-hyper}

\usepackage{bm}

\begin{document}

\preprint{AIP/123-QED}

\title{Unified 1D Theory and Design Principles for Harmonic Electrothermal Characterization of Nanoscale Conductors}

\author{Chuyue Peng}
\author{Joshua Ginzburg}
\author{Annika Shah}
\author{Matthias Kuehne}
\thanks{Author to whom correspondence should be addressed: kuehne@brown.edu}
\affiliation{%
Department of Physics, Brown University, Providence, Rhode Island 02912
}%

\date{\today}

\begin{abstract}
Electrothermal characterization based on the third or other harmonics of an ac Joule heating current is widely deployed for the thermal analysis of solid conductors and their environment, including solid substrates and fluids. However, a unified theory that bridges heat transfer in two archetypal experimental geometries - suspended vs. substrate-supported conductor - has been missing. Here, we present and validate such a theory that explicitly accounts for finite conductor length, thermal mass, and environmental coupling through a unified thermal transfer function. This framework enables the prediction of voltage responses at all harmonics of the driving current (dc, 1$\omega$, 2$\omega$, 3$\omega$) and the formulation of design principles for the characterization of nanoscale conductors. The conductor length $l$ is the primary parameter controlling the frequency regime at which the conductor's thermal mass dominates the thermal response, with the characteristic frequency $\omega_\mathrm{c}=\alpha/l^2$, where $\alpha$ is the conductor's thermal diffusivity - closely related to a criterion previously reported for suspended wires free from environmental coupling. Our unified framework generalizes this result, revealing that sufficiently weak environmental coupling is a necessary condition for $\omega_\mathrm{c}$ to govern the onset of thermal-mass-dominated response. Optimization of interfacial thermal resistance and environmental thermal impedance may further improve temperature resolution and facilitate on-substrate implementations.
\end{abstract}

\maketitle

Accurate measurement of thermal properties—thermal conductivity $\kappa$, specific heat $c_p$, and thermal diffusivity $\alpha$—is essential across diverse fields including materials science, microelectronics, and energy systems.\cite{Cahill2003,Cahill2014,Pop2010,Balandin2011,Marconnet2013} Among the various experimental techniques developed for thermal characterization, electrothermal methods using line heaters have emerged as particularly powerful approaches due to their use of narrow-band lock-in detection, which provides excellent signal-to-noise ratios.\cite{Cahill1990,Cahill1994,moon19963omega,Lee1997,Yi1999,Lu2001,Dames2005,Dames2013,Bhardwaj2022,Yang2025} In these techniques, an electrical conductor serves simultaneously as both heater and thermometer: an alternating current at frequency $\omega$ produces Joule heating at 2$\omega$, which generates temperature oscillations that modulate the heater resistance, ultimately yielding voltage signals at multiple harmonics (1$\omega$, 2$\omega$, 3$\omega$) that encode information about the thermal properties of the system.\cite{Dames2005}

Alongside electrothermal approaches, a variety of other techniques have been developed for the thermal characterization of nanoscale conductors. The thermal bridge method, in which a nanowire or nanotube is suspended between two microfabricated islands (one acting as heater and the other as thermometer) was among the first platforms to yield thermal conductance measurements of individual nanotubes and nanowires.\cite{Kim2001,Kim2002,Pettes2009} Suspended conductors have also been characterized by steady-state Joule heating coupled with electrical resistance readout\cite{Fujii2005,Pop2005} or scanning thermal microscopy.\cite{Kim2002,Shi2009} Optical thermometry based on Raman spectroscopy has further enabled the determination of thermal conductivity of nanotubes.\cite{Hsu2008,Kuehne2021} Electron thermal microscopy, in which phase transitions in nanoscale metal islands deposited on a membrane report the local temperature while the conductor is Joule-heated, has provided spatially resolved thermal information.\cite{Bergtrup2007,Brintlinger2008,Baloch2010,Baloch2012} Local electron-beam heating of a nanowire in thermal bridge configuration allows spatially resolved thermal resistance profiling and avoids systematic errors from contact resistances.\cite{Wang2011,Liu2014} A feature common to all of these approaches is that they probe steady-state thermal conditions and therefore provide access only to the thermal conductivity of the material under test. Harmonic electrothermal measurements, by contrast, encode both thermal conductivity and heat capacity in the frequency-dependent voltage response, making them uniquely suited for complete thermal characterization of nanoscale conductors. A unified theoretical framework that quantitatively informs the design and interpretation of such measurements is therefore of broad relevance, and constitutes the subject of the present work.

Despite widespread adoption, the theoretical frameworks developed for different experimental configurations have remained largely disconnected. For suspended wire geometries, one-dimensional heat conduction models explicitly incorporate the wire's thermal mass and finite length, enabling the extraction of both thermal conductivity and specific heat.\cite{Yi1999,Lu2001,Dames2005} In contrast, for line heaters deposited on solid substrates the standard analysis assumes uniform heating along the length of the heater, neglects the conductor's thermal mass, and ignores any interfacial thermal resistance (ITR) between the heater and the substrate. Instead, one usually only considers the reduced problem of heat diffusion into the substrate in the direction transverse to the heater axis.\cite{Cahill1990,Cahill1994,moon19963omega,Lee1997,Dames2013} A unified theoretical description that encompasses both configurations as limiting cases is currently missing from the literature. Such a framework would not only provide conceptual clarity but would also predict the frequency regimes in which the heat capacity of the conductor becomes significant and quantitatively inform the impact of ITR and environmental thermal impedance—information critical for experimental design and accurate data interpretation.

\begin{figure}
\includegraphics[width=0.45\textwidth]{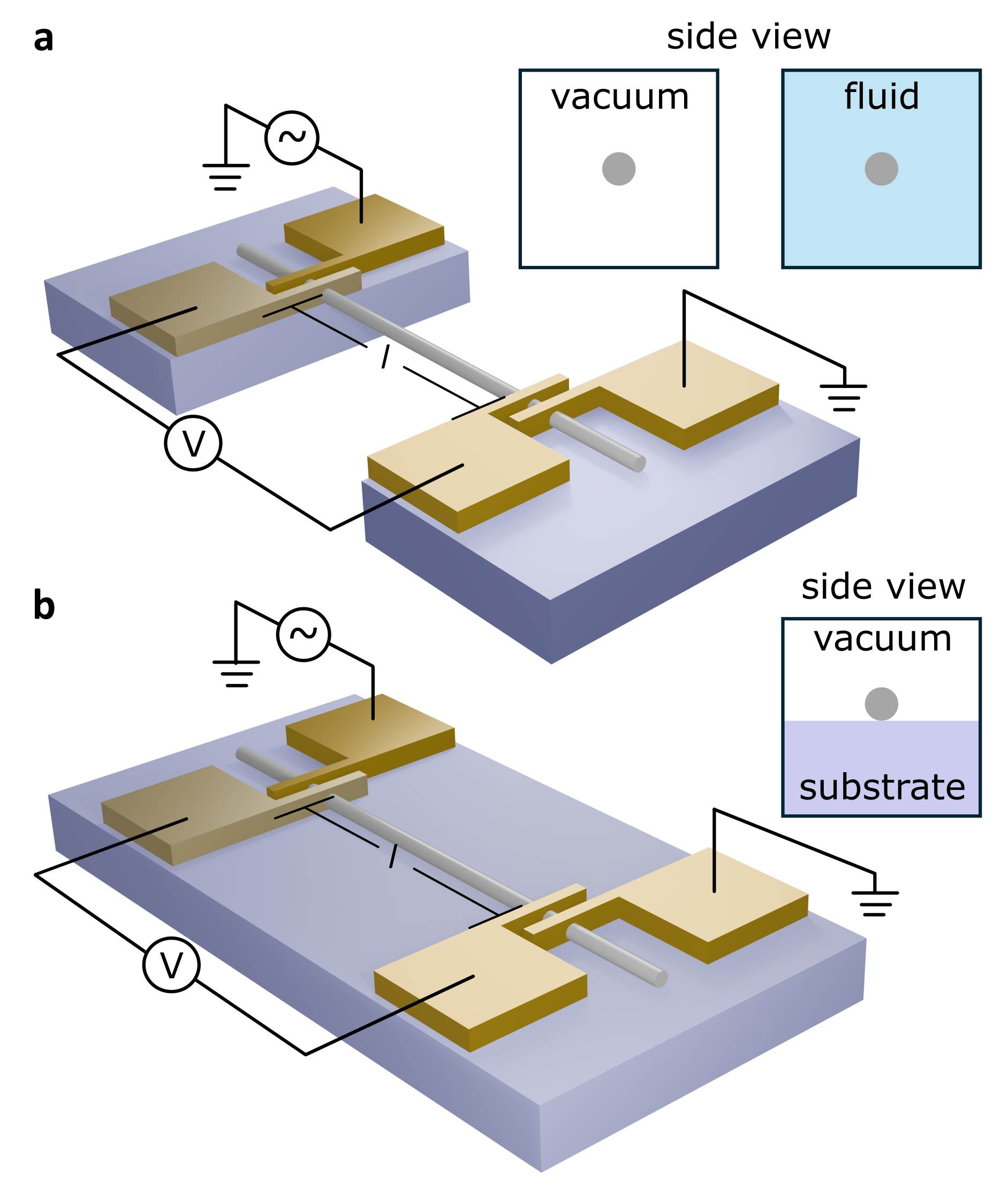}
\caption{\label{fig:1} Schematic geometries of harmonic electrothermal characterization for which we present a unified theory: (a) suspended conductor and (b) conductor on substrate. The space around the conductor could be vacuum or fluid.}
\end{figure}

Herein, we present a unified theory that addresses these limitations and provides quantitative design criteria for electrothermal characterization across different experimental configurations. Our framework predicts when the thermal mass of the conductor becomes significant and how environmental coupling can be optimized, providing critical information for experiment design and accurate data interpretation. We derive a unified thermal transfer function by solving the one-dimensional heat transfer equation that explicitly accounts for the finite length $l$ of the conductor illustrated in Fig.~\ref{fig:1}, its thermal mass $C=\rho c_plS$, and its coupling to different environments:
\begin{equation}
    \label{eq:1d_heat_transfer_g}
    \rho c_p\frac{\partial\theta(x,t)}{\partial t} = \kappa \frac{\partial^2\theta(x,t)}{\partial x^2} +\frac{Q(t)}{Sl}-\frac{hP}{S}\theta(x,t).
\end{equation}
Here, $\rho$ is the mass density of the heater, $\theta(x,t)$ is the temperature rise as a function of time $t$ and position $x$ along the heater, $Q(t)$ is the heat generated by the heater as a function of time $t$, $h$ is the effective heat transfer coefficient quantifying heat transfer to the environment in units of W/(m$^2$K), $l$ is the length of the conductor section between the two inner electrical contacts, $S$ is the heater cross-section, and $P$ is the part of the circumference in contact with the non-vacuum environment. Our framework suggests $l$ as the primary parameter to control the characteristic frequency $\omega_\mathrm{c}$ above which the conductor's thermal mass dominates at least as long as the effective thermal impedance of the environment $|Z_h|$ is kept larger than the conductor's axial thermal resistance $R_\mathrm{th}$. This further establishes quantitative criteria for optimizing environmental coupling through design of ITR and environmental thermal impedance.

In writing Eq.~\ref{eq:1d_heat_transfer_g}, we assume cross-sectionally uniform temperature in the conductor at every position $x$. We are generally concerned with maximum temperature rises on the order of 1~K and ignore thermal radiation effects discussed elsewhere.\cite{Lu2001,Dames2013,Peng2025} We further assume negligible temperature rise at the two ends of the conductor section between the two inner electrical contacts, i.e., $\theta(0,t)=\theta(l,t)=0$. We are further concerned with joule heating through driving an electrical current along the conductor (using the two outer electrical contacts) that generally may have both dc and ac components. It is convenient to solve Eq.~\ref{eq:1d_heat_transfer_g} in frequency domain where dc and ac responses can be separated\cite{Dames2005,Dames2013} assuming that $h$ is at most a function of $\omega$ but not a function of time $t$. Hence, the ac part in the frequency domain can be written as
\begin{equation}
    i\omega \rho c_p \theta_{\omega}(x) = \kappa \frac{\partial^2\theta_{\omega}(x)}{\partial x^2} + \frac{Q_\omega}{Sl} -\frac{hP}{S}\theta_\omega (x).
\end{equation}

The average (along the length $l$) temperature rise of the conductor $\theta_{\omega,\mathrm{ave}}$ can be related to the heat input $Q_\omega$ through the thermal transfer function $\mathcal{Z}$.\cite{Dames2005} In the frequency domain, this relation can be written as:
\begin{equation}
\label{eq:theta_wave}
    \theta_{\omega,\mathrm{ave}} = Q_{\omega}\mathcal{Z}(\omega).
\end{equation}
Assuming a linear response of the conductor's resistance to $\theta_{\omega,\mathrm{ave}}$, knowledge of the specific form of $\mathcal{Z}$ allows predicting the voltage response between the two inner electrical contacts at the various harmonics of the frequency $\omega$ of the applied current,\cite{Dames2005} see Sec.~I of the Supplementary Information. For example, the root-mean-square (rms) $3\omega$ voltage can be written as:
\begin{equation}
\label{eq:3w_by_impedance}
    V_{3\omega,\mathrm{rms}}=\frac{I_\mathrm{rms}^3RR'}{2}|\mathcal{Z}(2\omega)|,
\end{equation}
where $R$ is the resistance of the conductor, $R'$ is its temperature coefficient, and $I_\mathrm{rms}$ is the rms magnitude of the applied electrical current.

We derive the following unified thermal transfer function (Sec.~II of the Supplementary Information):
\begin{align}
\label{eq:thermal_transfer}
    \nonumber \mathcal{Z}(\omega) &= \\ &\frac{1}{Z_C^{-1}+Z_h^{-1}}\left[1-\frac{\tanh{\left(\sqrt{R_\mathrm{th}\left(Z_C^{-1}+Z_h^{-1}\right)}/2\right)}}{\sqrt{R_\mathrm{th}\left(Z_C^{-1}+Z_h^{-1}\right)}/2}\right].
\end{align}
Here, $Z_C=1/i\omega C$ is the conductor's thermal mass impedance, $Z_h=1/hPl$ is the effective thermal impedance of the environment, and $R_\mathrm{th}=l/\kappa S$ is the axial thermal resistance. $\mathcal{Z}(\omega)$ represents a parallel combination of $Z_C$ and $Z_h$ that is modulated by the roll-on term in square brackets. The roll-on term itself depends on the ratio of $R_\mathrm{th}/(Z_C^{-1}+Z_h^{-1})^{-1}$, which acts as a complex Biot-like number that characterizes how much heat is locally stored or transmitted to the environment vs. conducted away axially along the heater.

\begin{figure}
\includegraphics[width=0.49\textwidth]{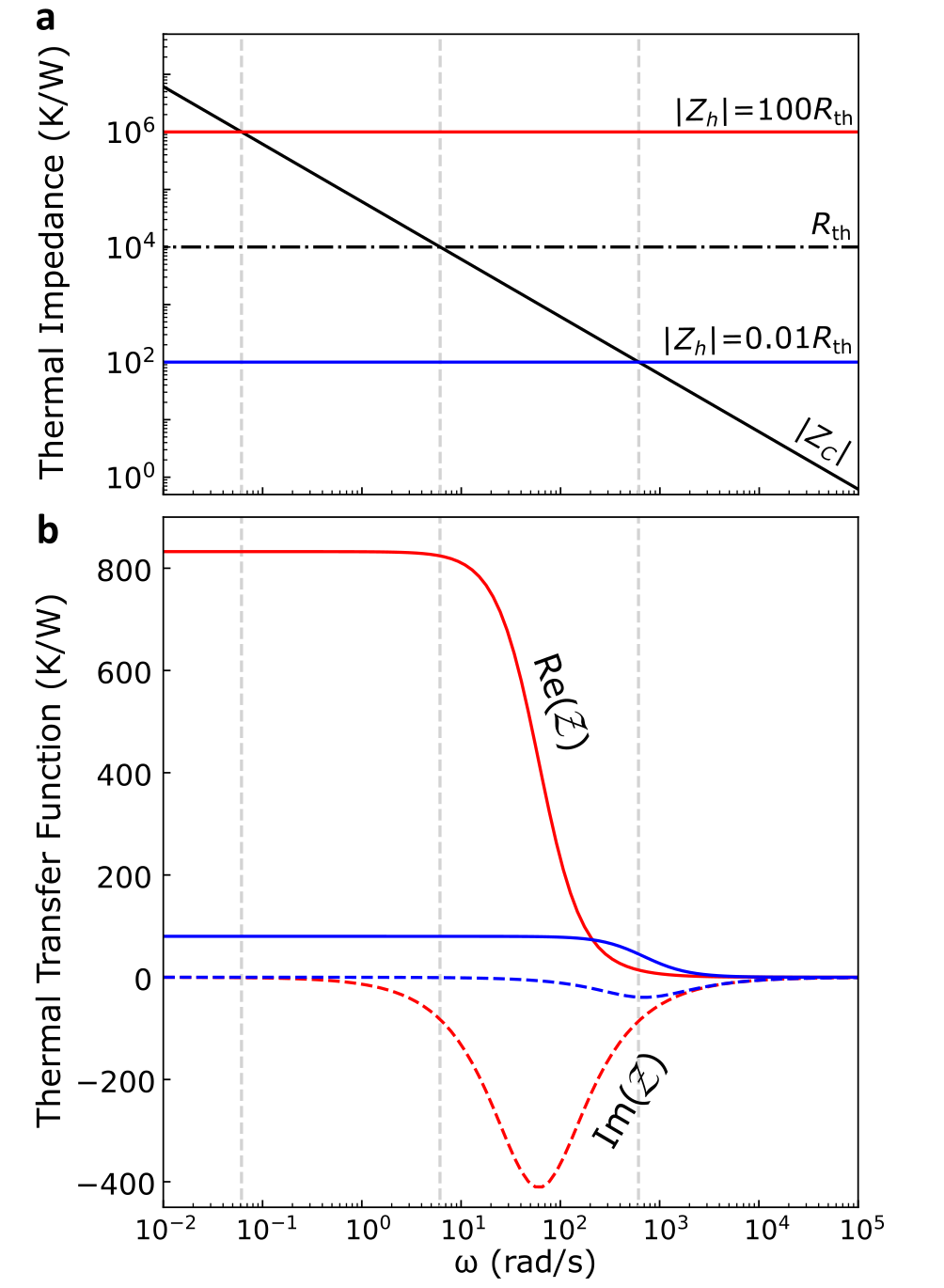}
\caption{\label{fig:2} Dissecting the unified thermal transfer function $\mathcal{Z}$, Eq.~\ref{eq:thermal_transfer}. (a) Magnitude of thermal impedances entering $\mathcal{Z}$. (b) Real (in-phase, solid line) and imaginary (out-of-phase, dashed line) parts of $\mathcal{Z}$. We compare two illustrative cases: $|Z_h|>R_\mathrm{th}$ (red) and $|Z_h|<R_\mathrm{th}$ (blue). Vertical dashed lines indicate frequencies where $|Z_C|=|Z_h|$ or $|Z_C|=R_\mathrm{th}$.}
\end{figure}

In Fig.~\ref{fig:2}, we schematically show how the different thermal impedances affect Eq.~\ref{eq:thermal_transfer}. Assuming fixed $R_\mathrm{th}$ and $C$, we consider two illustrative cases: $|Z_h|>R_\mathrm{th}$ and $|Z_h|<R_\mathrm{th}$. These two conditions represent different values of $h$, which here we consider independent of frequency. At low frequencies, the real (in-phase) component of $\mathcal{Z}$ assumes a constant value whereas the imaginary (out-of-phase) component approaches 0. As can be seen, when $|Z_h|>R_\mathrm{th}$ (red curves) the roll-off in $\mathrm{Re}(\mathcal{Z})$ is determined by the frequency at which $|Z_C|\approx R_{\mathrm{th}}$. On the other hand, when $|Z_h|<R_\mathrm{th}$ (blue curves) the roll-off in $\mathrm{Re}(\mathcal{Z})$ is determined by the frequency at which $|Z_C|\approx |Z_h|$. Note that at high enough frequencies where $|Z_C|<|Z_h|$ and $|Z_C|<R_\mathrm{th}$ Eq.~\ref{eq:thermal_transfer} is solely determined by the conductor thermal mass and assumes the simple form $\mathcal{Z}=Z_C=(i\omega C)^{-1}$.

While Eq.~\ref{eq:thermal_transfer} is the most general way of stating the thermal transfer function, a few comments are in order. First, for the important case of a free-standing wire in vacuum,\cite{Yi1999,Lu2001,Dames2005} Eq.~\ref{eq:thermal_transfer} simplifies to
\begin{equation}
\label{eq:zavevacuum}
    \mathcal{Z}_\mathrm{vac}(\omega)=Z_C\left[1-\frac{\tanh\left(\sqrt{R_\mathrm{th}/Z_C}/2\right)}{\sqrt{R_\mathrm{th}/Z_C}/2}\right].
\end{equation}
An equivalent expression has previously been derived in Ref.~(\onlinecite{Dames2005}). The low-frequency limit of Eq.~\ref{eq:zavevacuum} can be written as
\begin{equation}
    \lim_{\omega\rightarrow0} \mathcal{Z}_\mathrm{vac}(\omega)=\frac{l}{12 S\kappa},
\end{equation}
which is independent of both $\omega$ and $\rho c_p$ as expected.\cite{Lu2001}

\begin{figure}
\includegraphics[width=0.49\textwidth]{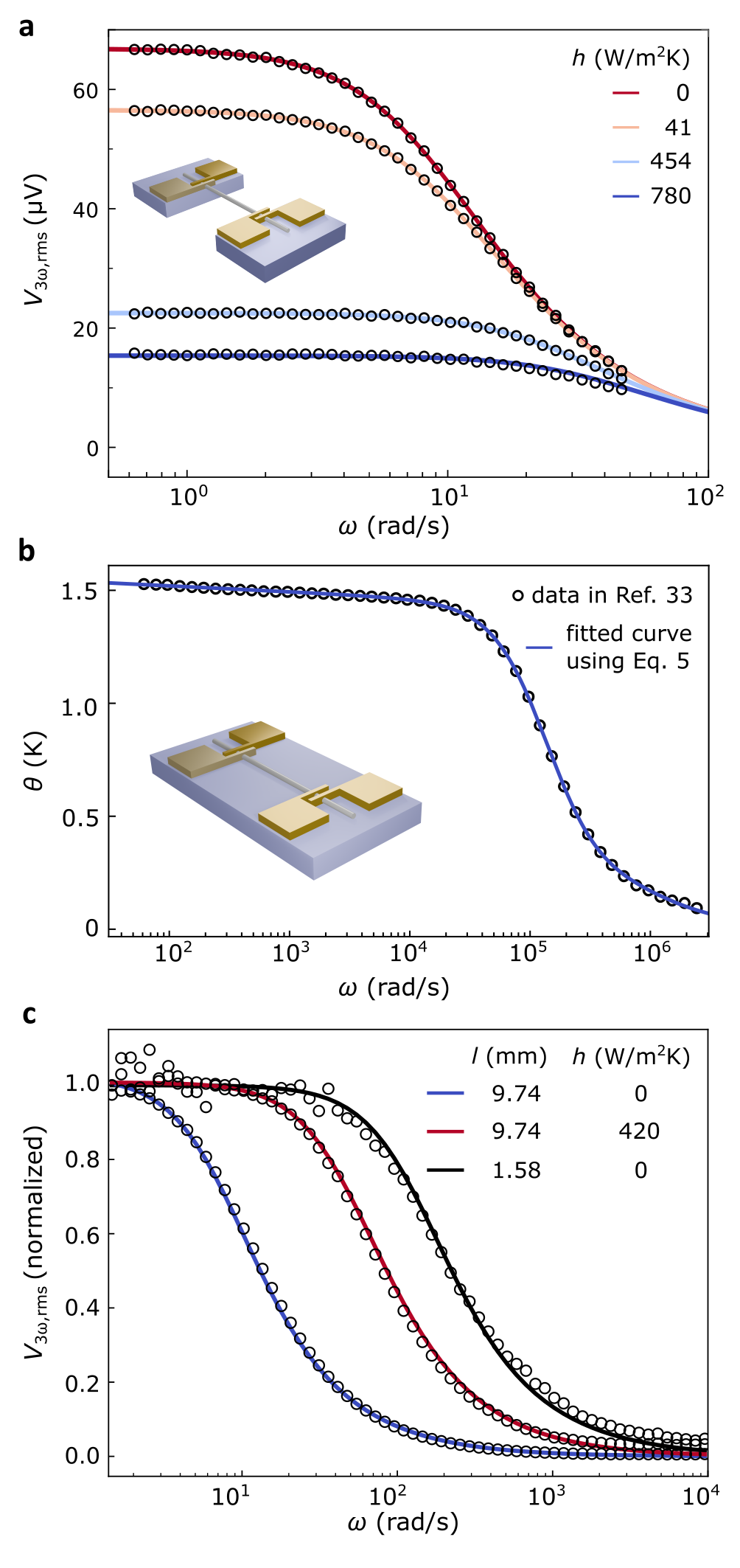}
\caption{\label{fig:3} Validating the unified thermal transfer function. (a) Best fit (solid lines) using Eqs.~\ref{eq:3w_by_impedance} and \ref{eq:thermal_transfer} of $V_{3\omega}$ data from Ref.~\onlinecite{Peng2025} (open circles) measured on a platinum wire suspended in vacuum (red) and different rarified gas atmospheres yielding the listed $h$ values, $\kappa=74.5\pm0.1$ ~W/(m$\cdot$K), and $\rho c_p=(3.13\pm0.01)\times10^6$~J/(m$^3$K). (b)Average temperature rise measured on AlSiCu line heaters on SiO$_2$-terminated silicon substrates reported in Ref.~\onlinecite{Raudzis2003} (open circles). The blue line shows a fitted curve using Eq.~\ref{eq:thermal_transfer}, yielding volumetric heat capacity of the heater $(\rho c_p)_h = (2.21\pm 0.02)\times10^6$ J~m$^{-3}$K$^{-1}$, thermal conductivity of the film $\kappa_{\mathrm{f}} = (0.996\pm0.002)$ W~m$^{-1}$K$^{-1}$ and thermal conductivity of the substrate $\kappa_{\mathrm{s}} = (147\pm3)$ W~m$^{-1}$K$^{-1}$ (see Sec.~III of the Supplementary Information for detailed discussion). Insets are schematic sample geometries as in Fig.~\ref{fig:1}. (c) Normalized $V_{3\omega}$ data of free-standing carbon nanotube wires with different lengths $l$ measured in high ($h=0$) and low-vacuum conditions.}
\end{figure}

Second, for the more general case of a wire in contact with a surrounding fluid, heat transfer may at least in some cases be accurately described using a frequency-independent $h$. We have recently demonstrated this for the case of platinum wire in rarified gas atmospheres.\cite{Peng2025} In Fig.~\ref{fig:3}a, we successfully re-analyze data from this recent work\cite{Peng2025} using Eqs.~\ref{eq:3w_by_impedance} and \ref{eq:thermal_transfer} yielding the frequency-independent $h$ values stated in the figure. The obtained values compare very well with those previously reported, which validates the unified thermal transfer function used here. Note that at least in the absence of strong convection, $h$ in this context of heat transfer to a surrounding fluid can be interpreted as an in-series combination of solid-fluid interfacial (Kapitza) resistance\cite{Swartz1989,Chen2022} and the fluid's thermal impedance. In the presence of significant convection, $h$ is more appropriately viewed as an effective parameter that lumps all heat loss mechanisms from the surface to the surrounding fluid. 

Third, Eq.~\ref{eq:thermal_transfer} also extends to the widely adopted case of a heater on a substrate and in vacuum.\cite{Cahill1990,Cahill1994,moon19963omega,Lee1997} In this case, $h^{-1}$ can be modeled as an in-series combination of ITR\cite{Swartz1989,Chen2022} and thermal impedance of the substrate, i.e., 
\begin{equation}
\label{eq:zh}
    Z_h(\omega)=\frac{1}{h_\mathrm{I}Pl}+Z_\mathrm{sub}(\omega).
\end{equation}
We emphasize that here $P$ should be interpreted as the width of the contact area between the conductor and the substrate. If, for the sake of simplicity, we consider an isotropic, semi-infinite substrate, we can write
\begin{equation}
\label{eq:zsub}
    Z_\mathrm{sub}(\omega)=\frac{1}{l\pi \kappa_\mathrm{sub}}\int_0^\infty\frac{\sin^2(kb)}{(kb)^2\sqrt{k^2+q^2}}dk,
\end{equation}
where the magnitude of $1/q=\sqrt{\alpha_\mathrm{sub}/i\omega}$ is the diffusive wavelength in the substrate, and $b$ is the half width of the heater,\cite{Cahill1990} i.e., $2b=P$. The description can be extended to the more general case where the substrate may have anisotropic thermal properties or may consist of multiple layers using models reported elsewhere (see also Sec.~III of the Supplementary Information).\cite{borca2001data,Tong2006,Dames2013} Note that if we assume negligible conductor thermal mass, i.e., $|Z_C^{-1}|\ll |Z_h^{-1}|$, Eq.~\ref{eq:thermal_transfer} simplifies to
\begin{equation}
    \mathcal{Z}(\omega)=Z_h(\omega)\left[1-\frac{\tanh\left(\sqrt{R_\mathrm{th}/Z_h(\omega)}/2\right)}{\sqrt{R_\mathrm{th}/Z_h(\omega)}/2}\right].
\end{equation}
If we further assume $R_\mathrm{th}^{-1}\ll|Z_h^{-1}|$, we recover $\mathcal{Z}(\omega)=Z_h(\omega)$. If ITR between heater and substrate is negligible, we recover $\mathcal{Z}(\omega)=Z_\mathrm{sub}$. This case was originally discussed by Cahill.\cite{Cahill1990} In Fig.~\ref{fig:3}b we compare the average temperature rise calculated for a strip heater on a silicon substrate terminated with 2~$\mathrm{\mu m}$ thick SiO$_2$ using Eq.~\ref{eq:thermal_transfer} with data from Ref.~\onlinecite{Raudzis2003}. For this calculation, we use $Z_h=Z_\mathrm{sub}$ (i.e., ITR neglected) and a thin-film-on-semi-infinite-substrate model for $Z_\mathrm{sub}$ reported in Ref.~\onlinecite{borca2001data} as well as parameters from Table 1 in Ref.~\onlinecite{Raudzis2003}.

Fourth, Eq.~\ref{eq:thermal_transfer} can in principle be further extended to the more general case of a wire on substrate and immersed in a fluid. This would entail replacing the third term on the right of Eq.~\ref{eq:1d_heat_transfer_g} with two independent terms of this form. Eq.~\ref{eq:thermal_transfer} can in principle also be applied to a 2D material\cite{Gu2018} conductor of length $l$, width $P$, and thickness $S/P$ as an extension of its demonstrated applicability to measurements in supported strip heater configuration (Fig.~3b). Limitations might arise from possible spatial variations in Joule heating that would violate the cross-sectionally uniform temperature assumed by Eq.~\ref{eq:1d_heat_transfer_g}.\cite{Bao2011} For the case of nanoscale conductors on polar substrates another limitation could arise from remote Joule heating, i.e., the direct dissipation of electrical power from the conductor via near-field remote scattering of hot electrons off surface polaritons in the substrate.\cite{Baloch2012,Petrov2006,Rotkin2009}

The preceding four points serve to highlight the generality of Eq.~\ref{eq:thermal_transfer} and how this equation reduces to more specialized expressions relevant to common implementations in the field. Fig.~\ref{fig:3}a and b shows that the unified thermal transfer function accurately describes measurements from both archetypal geometries shown in Fig.~\ref{fig:1}. In the remainder of this letter, we turn to deriving design principles for electrothermal measurements based on Eq.~\ref{eq:thermal_transfer}. A particular question we seek to answer is how to design an experiment geared towards measuring $C$ of a heater with nanoscale cross-section such as is relevant for the thermal characterization of nanowires or nanotubes.\cite{Lu2001,Marconnet2013}

\begin{figure}
\includegraphics[width=0.49\textwidth]{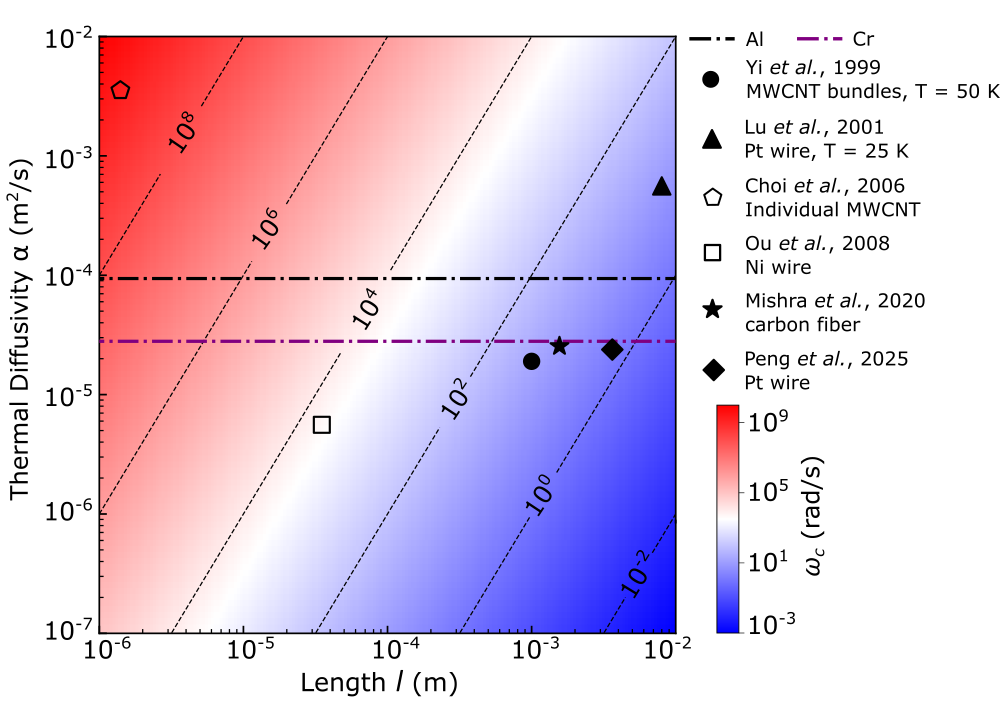}
\caption{\label{fig:4} Characteristic frequency $\omega_\mathrm{c}$, Eq.~\ref{eq:constraint1}, above which $|Z_C|<R_\mathrm{th}$. Markers locate various conductors with micro- to nanoscale cross-section studied in the literature.\cite{Yi1999,Lu2001,Choi2006,ou2008electrical,Mishra2020,Peng2025} Dash-dotted lines indicate room-temperature thermal diffusivities of select metals for comparison.\cite{Powell1966}}
\end{figure}

In designing an experiment geared towards measuring $C$, an important consideration is the frequency range at which $\mathcal{Z}$ is dominated by $Z_C^{-1}$. For $Z_C^{-1}$ to dominate, we need $|Z_C|<|R_\mathrm{th}|$ as well as $|Z_C|<|Z_h|$ (see Fig.~\ref{fig:2}). $R_\mathrm{th}$ is set by $\kappa$ as well as the conductor dimensions $l$ and $S$. Since $C\propto lS$, we conclude that adjusting $l$ is of primary importance in setting the frequency above which $Z_C^{-1}$ dominates, while changing $S$ is not. In Fig.~3c we present normalized $V_{3\omega,\mathrm{rms}}$ data measured on carbon nanotube wires (Nanografi) of different lengths (see Sec.~IV of the Supplementary Information for details). As can be seen, increasing the wire length from 1.58~mm to 9.74~mm downshifts the roll-off by over one decade when $h=0$. This roll-off is set by the frequency at which $R_\mathrm{th}=|Z_C|$ (crossing of the two black lines in Fig.~\ref{fig:2}a):
\begin{equation}
\label{eq:constraint1}
\omega_\mathrm{c}=\frac{\alpha}{l^2}.
\end{equation}
Here, $\alpha=\kappa/\rho c_p$ is the thermal diffusivity of the conductor. For the short and long wire data at $h=0$ we get $\omega_\mathrm{c,short}=(26\pm10)$~rad/s and $\omega_\mathrm{c,long}=(1.5\pm0.2)$~rad/s, respectively. In Fig.~\ref{fig:4} we plot $\omega_\mathrm{c}$ as a function of both $\alpha$ and $l$. Markers situate previous studies on the parameter space. Solid markers indicate that $\alpha$ and $l$ were reported, while hollow markers indicate that we used an estimate of $\rho c_p$ to calculate $\alpha$. A criterion similar to Eq.~\ref{eq:constraint1} (up to a factor of $\pi^2$) has been used in the study of suspended conductors free from environmental coupling only.\cite{Lu2001}

A second characteristic frequency is given by $|Z_C|=|Z_h|$ as
\begin{equation}
\label{eq:constraint2}
\omega_\mathrm{c}'=\frac{|h|Pl}{C}=\frac{|h|P}{\rho c_pS},
\end{equation}
which is independent of the conductor length $l$. A criterion similar to Eq.~\ref{eq:constraint2} can be derived from a transverse heat transfer model discussed in Sec.~V of the Supplementary Information. For a conductor with circular cross-section $S=\pi r^2$ thermally coupled to the environment along its entire circumference $P=2\pi r$, Eq.~\ref{eq:constraint2} would further simplify to $\omega_\mathrm{c}'=2|h|/\rho c_pr$. This unveils a potential challenge in the characterization of conductors with nanoscale cross-section as small $r$ acts to impose a large $\omega_\mathrm{c}'$. However, $h$ is a parameter that may in principle be controlled by experiment design. Since we can always in principle resort to the geometry of a free-standing conductor in vacuum for which $Z_h\rightarrow\infty$, but we cannot arbitrarily increase $R_\mathrm{th}$, adjusting $R_\mathrm{th}$ through $l$ should indeed be the primary knob for setting the frequency range, ideally keeping $\omega_\mathrm{c}\ge\omega_\mathrm{c}'$. However, we are still allowed to choose $|Z_h|\ge|R_\mathrm{th}|$ without affecting the frequency range, see Fig.~\ref{fig:2}. That is,
\begin{equation}
\label{eq:constraint3}
|h_\mathrm{max}|=\frac{\kappa S}{Pl^2}   
\end{equation}
would be the maximum allowable $h$. Again, for a conductor with circular cross-section $S=\pi r^2$ and $P=2\pi r$, we would have $|h_\mathrm{max}|=\kappa r/2l^2$. In Fig.~3c, the red curve is a best fit of $V_{3\omega}$ data measured on the $l = 9.74$~mm carbon nanotube wire in low-vacuum conditions yielding $h=(420\pm 30)~\mathrm{W/(m^2K)}$ and through use of Eq.~\ref{eq:constraint2} $\omega_\mathrm{c,long}'=(80\pm10)$~rad/s which is larger than $\omega_\mathrm{c,long}$. Compared with the high-vacuum measurement, the roll-off has shifted back to higher frequencies, partially canceling the frequency reduction achieved by designing a longer wire sample in the first place. This illustrates the tradeoff in design criteria provided Eqs.~\ref{eq:constraint1} and \ref{eq:constraint2}, with a constraint imposed by Eq.~\ref{eq:constraint3} which here yields $h_\mathrm{max} = (8\pm1)~\mathrm{W/(m^2K)}$ for the long carbon nanotube wire. Allowing for a finite $|Z_h|$ may, however, have several benefits. For example, finite $h$ makes the temperature rise along the length of the conductor more uniform, which increases temperature resolution. Finite $h$, if set by an underlying substrate, may improve device fabrication workflow since making a conductor on a suitable substrate may be more feasible in some cases than having it free-standing. Note that although finite $h$ acts to suppress the voltage response, the $I_\mathrm{rms}$ may be increased to increase signal strength while keeping the average temperature rise fixed.

For a conductor on a substrate, note that from Eq.~\ref{eq:zh} $|Z_h|\ge1/h_\mathrm{I}Pl$ equivalently $|h|\le h_\mathrm{I}$. In other words, the ITR sets a fundamental constraint on experiment design because $|h_\mathrm{max}|\le h_\mathrm{I}$. Typical values of $h_\mathrm{I}$ between dissimilar solids can span several orders of magnitude, ranging from tens of $\mathrm{MW/m^2K}$ to tens of $\mathrm{GW/m^2K}$.\cite{Chen2022} For example, for carbon nanotubes on SiO$_2$ typical ITR values are on the order of $10^{-8}$~m$^2$K/W.\cite{Pop2007,Ong2010,Liao2010} For a given conductor, the ITR may be increased by choosing a substrate that minimizes interfacial bonding or maximizes the vibrational mismatch such as through selecting materials with vastly different phonon density of states or by selecting weak van der Waals bonding over covalent bonding. For special cases such as nanoscale conductors with circular cross-section also analytical models of $h_\mathrm{I}$ have been put forward that may inform the engineering of environmental coupling.\cite{Bahadur2005,Prasher2005}

We have developed a unified theoretical framework for harmonic electrothermal characterization of conductors that includes their thermal mass and environmental coupling. The unified thermal transfer function derived in Eq.~\ref{eq:thermal_transfer} enables prediction of voltage responses at all harmonics (dc, 1$\omega$, 2$\omega$, 3$\omega$) and encompasses suspended, fluid-immersed, and substrate-supported configurations. Our analysis establishes that conductor length $l$ should be the primary design parameter to control the frequency regime at which thermal mass dominates the response, with the characteristic frequency given by $\omega_\mathrm{c}=\alpha/l^2$. Additionally, careful design of thermal coupling to the environment, e.g., through control of ITR, may enable optimization of measurement sensitivity while allowing practical fabrication strategies. This framework thus provides quantitative design criteria for challenging nanoscale thermal characterization, establishing optimal combinations of conductor geometry, environmental coupling, and measurement frequency.

\section*{Supplementary Material}
I. From thermal transfer function to voltage response; II. Solution to the heat transfer equation; III. Multilayered substrates including with anisotropic thermal properties; IV. 3$\omega$ measurement on carbon nanotube wires; V. Transverse heat transfer model including conductor thermal mass and ITR.

\begin{acknowledgments}
This material is based upon work supported by the National Science Foundation under Grant No. 2341781. J. G. acknowledges support from the Brown University Summer/Semester Projects for Research, Internship, and Teaching (SPRINT) program through both an Undergraduate Teaching and Research Awards (UTRA) fellowship as well as an Advanced Undergraduate Research Fellowship. We thank John Cumings, Jacob Bair, and Jiaxuan Chen for their valuable input.
\end{acknowledgments}

\section*{Author contributions}
\textbf{Chuyue Peng}: Data curation, Formal analysis, Investigation, Methodology, Validation, Visualization, Writing - original draft, Writing - review \& editing. \textbf{Joshua Ginzburg}: Formal analysis, Writing - review \& editing. \textbf{Annika Shah}: Formal analysis, Writing - review \& editing. \textbf{Matthias Kuehne}: Conceptualization, Formal Analysis, Funding acquisition, Methodology, Project administration, Supervision, Writing - original draft, Writing - review \& editing.

\section*{Data Availability Statement}
The data that support the findings of this study are openly available in Zenodo, at https://doi.org/10.5281/zenodo.20635443, reference number \onlinecite{Peng2026data}.

\section*{References}

\bibliography{Unified3w}

\end{document}


\preprint{AIP/123-QED}

\title{Supplementary Information for ``Unified 1D Theory and Design Principles for Harmonic Electrothermal Characterization of Nanoscale Conductors''}

\author{Chuyue Peng}
\author{Joshua Ginzburg}
\author{Annika Shah}
\author{Matthias Kuehne}
\thanks{Author to whom correspondence should be addressed: kuehne@brown.edu}
\affiliation{%
Department of Physics, Brown University, Providence, Rhode Island 02912
}%

\date{\today}

\maketitle


\section{From thermal transfer function to voltage response}
\label{app: voltage response}
We are interested in the voltage response of the heater at different harmonic frequencies. In a general case which was discussed by \textcite{Dames2005}, the current has the form:
\begin{equation}
    I(t) = I_{\mathrm{dc}} + I_{1}\sin(\omega_1 t).
\end{equation}
Following their notation $\eta = I_{\mathrm{dc}}/I_{1}$, we provide the dc, 1$\omega$, 2$\omega$ and 3$\omega$ voltage responses which was firstly derived by \textcite{Dames2005} as table~\ref{tab:table1} for reference to readers.
\begin{table*}[htbp]
    \centering
    \begin{tabular}{c|c|c}
    \hline
    \hline
        Harmonics & In-phase Electrical Transfer Function & Out-of-phase Electrical Transfer Function\\
        \hline
        0 & 0 & $\eta\sqrt{2}\{\frac{1}{2R'I^2_{1,\mathrm{rms}}}+[\eta^2 + 1/2]Z(0) +\mathrm{Re}[Z(\omega_1)]\}$\\
        1 & $\frac{1}{2R'I^2_{1,\mathrm{rms}}}+[\eta^2 +1/2]Z(0)+2\eta^2\mathrm{Re}[Z(\omega_1)]+(1/4)\mathrm{Re[Z(2\omega_1)]}$ &$(1/4)\mathrm{Im}[Z(2\omega_1)]+2\eta^2\mathrm{Im[Z(\omega_1)]}$\\
        2 & $\eta(1/2)\{\mathrm{Im}[Z(2\omega_1)]+2\mathrm{Im}[Z(\omega_1)]\}$ & $-\eta(1/2)\{\mathrm{Re}[Z(2\omega_1)]+2\mathrm{Re}[Z(\omega_1)]\}$\\
        3 & $-(1/4)\mathrm{Re}[Z(2\omega_1)]$ & $-(1/4)\mathrm{Im}[Z(2\omega_1)]$\\
    \hline
    \hline
    \end{tabular}
    \caption{The dc, $1\omega$, 2$\omega$ and 3$\omega$ voltage responses (electrical transfer functions) defined by the thermal transfer function in Eq.~(5) in the main text. Table adopted from \cite{Dames2005}.}
    \label{tab:table1}
\end{table*}
With the electrical transfer function provided, the voltage response can be written as
\begin{align}
    \nonumber V(t) = RR'I_1^3\sum_{n=0}^3 [ & X_n(\omega_1,\eta)\sin(n\omega_1 t) \\ & + Y_n(\omega_1,\eta)\cos(n\omega_1 t)],
\end{align}
or with the rms values:
\begin{equation}
    \frac{V_{n\omega,\mathrm{rms}}}{2RR'I_{1,\mathrm{rms}}^3} = X_n(\omega_1, \eta) +iY_n(\omega_1,\eta).
\end{equation}

For a typical 3$\omega$ measurement, we consider the heating current in the form
\begin{equation}
    I = I_1 \sin{\omega_1 t},
\end{equation}
and the heating power is
\begin{equation}
    Q(t) = \frac{I_1^2 R}{2}[1-\cos(2\omega_1 t)].
\end{equation}
The wire response can be expressed as
\begin{equation}
    \theta_{\omega,\mathrm{ave}}(t) = e^{2i\omega_1 t}\theta_{\omega,\mathrm{ave}}(2\omega_1).
\end{equation}
The heater has an electric resistance $R(T)$ which is a function of temperature $T$, and to measure the $3\omega$ signal it requires a nonzero temperature coefficient, or equivalently $R' = \mathrm{d}R/\mathrm{d}T \neq 0$. The heating current results in the temperature fluctuation on the heater
\begin{align}
    \nonumber R(t) &= R+R'\theta_{\mathrm{ave}}(t)  \\  &= R+R'[\theta_{\mathrm{DC,ave}}+e^{2i\omega_1 t}\theta_{\omega,\mathrm{ave}}(2\omega_1)].
\end{align}
Only the last term contributes to the $3\omega$ voltage:
\begin{align}
    \nonumber &I_0\sin{\omega t} \times R'\theta_{\omega,\mathrm{ave}}(2\omega)e^{2i\omega t} \\ = & \frac{I_0 R'}{2}|\theta_{\omega,\mathrm{ave}}(2\omega)|e^{i\phi}\times(-ie^{3i\omega t}+ie^{i\omega t}),
\end{align}
and the $3\omega$ voltage is:
\begin{equation}
    V_{3\omega}(t) = \frac{I_1 R'}{2}|\theta_{\omega,\mathrm{ave}}(2\omega_1)|\times[-ie^{i(3\omega_1 t +\phi)}],
\end{equation}
so the amplitude is
\begin{equation}
    V_{3\omega} = \frac{I_1 R'}{2}|\theta_{\omega,\mathrm{ave}}(2\omega_1)|.
\end{equation}
From here we get Eq.~(4) in the main text.
Note that we already used Eq.~(\ref{eq:q_omega}) and
\begin{equation}
    Q_{\omega}(2\omega) = -\frac{1}{2}I_1^2R.
\end{equation}

\section{Solution to the Heat Transfer Equation}
\label{app: derivation}

We start with the one-dimensional heat-transfer equation for a line heater of length $l$:
\begin{equation}
    \label{eq:1D}
    \rho c_p\frac{\partial\theta(x,t)}{\partial t} = \kappa \frac{\partial^2\theta(x,t)}{\partial x^2} +\frac{Q(t)}{Sl},
\end{equation}
Here at first we consider the vacuum case and suspended wire, thus no heat transfer to the surrounding environment is considered. The heater lies between $x = 0$ and $x = l$, thus the boundary conditions are:
\begin{equation}
    \frac{\partial \theta(\frac{l}{2},t)}{\partial x} = 0,\ \theta(0,t) = \theta(l,t) = 0.
\end{equation}
The temperature rise and the heat generated by the heater could be separated into dc and ac part:
\begin{equation}
    \theta(x,t) = \theta_{\mathrm{dc}}(x)+\theta_{\mathrm{ac}}(x,t), \ Q(t) = Q_{\mathrm{dc}}+Q_{\mathrm{ac}}(t).
\end{equation}
We assume
\begin{equation}
    \label{eq:q_omega}
    \theta_{\mathrm{ac}}(x,t) = \theta_{\omega}(x)e^{i\omega t}, \ Q_{\mathrm{ac}}(t) = Q_{\omega}e^{i\omega t}.
\end{equation}
Eq.~(\ref{eq:1D}) could be separated into an ac part
\begin{equation}
    \label{eq:1Dac}
    i\omega \rho c_p \theta_{\omega}(x)e^{i\omega t} = \kappa \frac{\partial^2\theta_{\omega}(x)}{\partial x^2}e^{i\omega t} + \frac{Q_\omega}{Sl}e^{i\omega t}
\end{equation}
and a dc part
\begin{equation}
    \kappa \frac{\partial^2\theta_{\mathrm{dc}}(x)}{\partial x^2}+\frac{Q_{\mathrm{dc}}}{Sl} = 0.
\end{equation}
The solution to the dc part reads:
\begin{equation}
    \theta_{\mathrm{dc}}(x) = \frac{Q_{\mathrm{dc}}x(l-x)}{2\kappa lS}.
\end{equation}
To solve the ac part, we define
\begin{equation}
    W(x,\omega) = \theta_\omega(x)-\frac{Q_\omega}{i\omega C},
\end{equation}
where $C = \rho c_p Sl$ is the thermal mass. Substitute for $\theta_\omega$ for the ac part of the heat-transfer equation [Eq.~(\ref{eq:1Dac})], we get
\begin{equation}
\label{eq:proper}
    \frac{i\omega}{\alpha}W(x,\omega) = \frac{\partial ^2}{\partial x^2}W(x,\omega),
\end{equation}
where $\alpha = \kappa/\rho c_p$ is the thermal diffusivity. The transformed boundary conditions are:
\begin{equation}
    W(0,\omega) = W(l,\omega) = -\frac{Q_\omega}{i\omega C}, \ \frac{\partial W(\frac{l}{2},\omega)}{\partial x} = 0.
\end{equation}
The general solution writes:
\begin{equation}
    W(x,\omega) = -\frac{Q_\omega}{i\omega C}\times \frac{e^{\sqrt{{i\omega}/{\alpha}}\cdot(l-x)}+e^{\sqrt{{i\omega}/{\alpha}}\cdot x}}{1+e^{\sqrt{{i\omega}/{\alpha}}\cdot l}},
\end{equation}
The expression for $\theta_\omega(x)$ reads:
\begin{equation}
    \label{eq:theta_omega1}
    \theta_\omega(x) = \frac{Q_\omega}{i\omega C}\bigg[ 1-\frac{\cosh{\big[\sqrt{i\omega/\alpha}(x-{l}/{2})\big]}}{\cosh{\big(\sqrt{i\omega/\alpha}\times{l}/{2}\big)}} \bigg].
\end{equation}

When the coupling between the wire and different environment is considered, the proper solution should be modified:
\begin{equation}
    W(x,\theta) = \theta_\omega(x)-\frac{Q_\omega}{\kappa Sl(i\omega/\alpha + hP/\kappa S)},
\end{equation}
and we get
\begin{equation}
    \bigg(\frac{i\omega}{\alpha}+\frac{hP}{\kappa S}\bigg)W = \frac{\partial ^2 W}{\partial x^2}
\end{equation}
with the transformed boundary condition
\begin{align}
      \frac{\partial W(\frac{l}{2},\omega)}{\partial x} &= 0, \nonumber\\  W(0,\omega) = W(l,\omega) &= -\frac{Q_\omega}{\kappa Sl(i\omega/\alpha + hP/\kappa S)}.
\end{align}
The solution is
\begin{equation}
    W(x,\omega) = -\frac{Q_\omega}{\kappa Sl(i\omega/\alpha + hP/\kappa S)}\times \frac{e^{q'\cdot(l-x)}+e^{q'\cdot x}}{1+e^{q'\cdot l}},
\end{equation}
where $q' = \sqrt{i\omega/\alpha + hP/\kappa S}$. The expression for $\theta(\omega)$ reads:
\begin{equation}
    \label{eq:theta_omega2}
    \theta_\omega(x) = \frac{Q_\omega}{i\omega C + hPl}\bigg[ 1-\frac{\cosh{\big[q'(x-{l}/{2})\big]}}{\cosh{\big(q'{l}/{2}\big)}} \bigg].
\end{equation}
Note that this expression is similar to Eq.~(\ref{eq:theta_omega1}), with only $\sqrt{i\omega/\alpha}$ replaced by $q'$, and $i\omega C$ in the denominator replaced by $(i\omega C + hPl)$.

We are interested in the average temperature rise over the full length of the heater. Taking the spatial average of $\theta_\omega(x)$,we get
\begin{equation}
    \label{eq:theta_ave}
    \theta_{\omega,\mathrm{ave}} = \frac{1}{l}\int_0^l{\theta_\omega (x) \mathrm{d}x} = \frac{Q_\omega}{i\omega C+hPl}\bigg[1-\frac{2\tanh{(q'l/2)}}{q'l}\bigg],
\end{equation}
which finally gives the thermal transfer function
\begin{equation}
\label{eq:impedance}
    \mathcal{Z}(\omega) = \frac{1}{i\omega C+hPl}\bigg[1-\frac{2\tanh{(q'l/2)}}{q'l}\bigg],
\end{equation}
where $q' = \sqrt{i\omega/\alpha+hP/\kappa S}$. This can also be written in the following form:
\begin{align}
\label{eq:thermal_transfer2}
    \nonumber \mathcal{Z}(\omega) &= \\ &\frac{1}{i\omega C+hPl}\bigg[1-\frac{\tanh{\left(\sqrt{l/\kappa S\cdot(i\omega C +hPl)}/2\right)}}{\sqrt{l/\kappa S\cdot(i\omega C +hPl)}/2}\bigg],
\end{align}
which is equivalent to Eq.~(5) in the main text.
Note that the solution is also valid for dc part if one takes $\omega = 0$.

Now we provide the amplitude and real part of Eq.~(\ref{eq:thermal_transfer2}).
For convenience, we define 
\begin{equation}
    k^2 = \sqrt{\left(\frac{\omega}{\alpha}\right)^2 +\left(\frac{hP}{\kappa S}\right)^2}, \phi = \frac{1}{2}\mathrm{Arg}\left(\frac{hP}{\kappa S}+i\frac{\omega}{\alpha}\right),
\end{equation}
and the amplitude of the thermal transfer function is
\begin{widetext}
    \begin{align}
        \nonumber |\mathcal{Z}(\omega)| = & \frac{1}{lk\sqrt{C^2\omega^2+h^2l^2P^2}[\cos(lk\sin\phi)+\cosh(lk\cos\phi)]}\times \\ \nonumber &\bigg[ \big[lk[\cos(lk\sin\phi)+\cosh(lk\cos\phi)]-2\sin\phi\sin(lk\sin\phi)-2\cos\phi\sinh(lk\cos\phi)\big]^{2} \\ & +4\big[\cos\phi\sin(lk\sin\phi)-\sin\phi\sinh(lk\cos\phi)\big]^2\bigg]^{1/2}
    \end{align}
\end{widetext}
The real part of the thermal transfer function could be written as
\begin{widetext}
    \begin{align}
         \mathrm{Re}[\mathcal{Z}(\omega)] =  lhP+\frac{-2(C\omega \cos{\phi+hlP\sin\phi})\sinh{(lk\cos\phi)}+2(C\omega\sin\phi-hlP\cos\phi)\sinh{(kl\cos\phi)}}{lk(C^2\omega^2+h^2 l^2 P^2)[\cos{(lk\sin\phi)+\cosh(lk\cos\phi)}]}.
    \end{align}
\end{widetext}

\section{Multilayered substrates including with anisotropic thermal properties}

As mentioned in the main text, in considering the case of a heater on a substrate and in vacuum, $Z_\mathrm{sub}$ can be adjusted to describe substrates that consist of multiple layers or have anisotropic thermal properties. Borca-Tasciuc et al.\cite{borca2001data} proposed a two-dimensional heat-conduction model to describe a substrate composed of $n$ layers, each with distinct in-plane and cross-plane thermal conductivities. The thermal impedance of such a substrate is given by

\begin{equation}
\label{eq:multilayer z_sub}
Z_\mathrm{sub} = -\frac{1}{l \pi \kappa_{y_1}} \int_{0}^{\infty} \frac{1}{A_1 B_1} \frac{\sin^2(kb)}{k^2 b^2}dk
\end{equation}
with
\begin{equation}
A_{i-1} = \frac{A_i \frac{\kappa_{y_i }B_i}
{\kappa_{y_{i-1}} B_{i-1}} \tan (B_{i-1} d_{i-1})  }
{ 1- A_i \frac{\kappa_{y_i}B_i}
{\kappa_{y_{i-1}} B_{i-1}} \tan (B_{i-1} d_{i-1}) },
\nonumber
\end{equation}
and 
\begin{equation}
B_{i} = \left(\frac{\kappa_{x_i}}{\kappa_{y_i}} k ^2 
+ \frac{i2\omega}{\alpha_{y_i}}\right)^{1/2}.
\nonumber
\end{equation}

Here, the index $i$ denotes the $i$th layer counting from the top of the substrate, $\kappa_x$ and $\kappa_y$ are in-plane and cross-plane thermal conductivities, respectively, $d$ is layer thickness, and $\alpha_{y}$ is the cross-plane thermal diffusivity. To compute $Z_\mathrm{sub}$, $B_1$ can be determined directly, but $A_1$ must be calculated iteratively starting from the bottom layer of the substrate. If the bottom layer is assumed to be semi-infinite, $A_n=-1$. If it is finite, the bottom boundary can be assumed to be adiabatic, in which case $A_n = -\tanh (B_n d_n)$, or isothermal, in which case $A_n = -1/\tanh (B_n d_n)$.

For the curve shown in Fig. 3b, parameters were taken from Raudzis et al.~\cite{Raudzis2003}, who describe a sample consisting of a 525 \textmu m Si bulk substrate, a 2.08 \textmu m SiO$_2$ thin film, and an AlSiCu strip heater which is 2.5 mm long and 14.4 \textmu m wide. The dimensions and thermal properties of this sample can be found in Table 1 of Ref.~\onlinecite{Raudzis2003}. When implementing Eq.~\ref{eq:multilayer z_sub}, the Si bulk substrate was assumed to be semi-infinite, hence giving $A_2 = -1$. In addition, both layers were assumed to have isotropic thermal properties, so the simplification $\frac{\kappa_{x_i}}{\kappa_{y_i}} = 1$ was made.

In Fig. 3b, the experimental data from Ref.~\onlinecite{Raudzis2003} are fitted to Eq.~5 in the main text using $Z_{\mathrm{sub}}$ calculated with the method described above. The fitting parameters include the volumetric heat capacity of the heater $(\rho c_p)_h$, the thermal conductivity of the film $\kappa_{\mathrm{f}}$ and thermal conductivity of the substrate $\kappa_{\mathrm{s}}$. In the fitting, the thermal conductivity of the heater is fixed at 150~W~m$^{-1}$K$^{-1}$, and the heating power is estimated as 0.0263~W. The volumetric heat capacity of the film $(\rho c_p)_{\mathrm{f}} = 1.7\times 10^6$~J~m$^{-3}$K$^{-1}$ and the substrate $(\rho c_p)_{\mathrm{s}} = 1.7\times 10^6$~J~m$^{-3}$K$^{-1}$ have also been used. Our fitting yields $(\rho c_p)_h = (2.21\pm 0.02)\times10^6$ J~m$^{-3}$K$^{-1}$, $\kappa_{\mathrm{f}} = (0.996\pm0.002)$ W~m$^{-1}$K$^{-1}$ and  $\kappa_{\mathrm{s}} = (147\pm3)$ W~m$^{-1}$K$^{-1}$. These values are consistent with previously reported results, $(\rho c_p)_h = (2.37\pm 0.19)\times10^6$ J~m$^{-3}$K$^{-1}$, $\kappa_{\mathrm{f}} = (1.02\pm0.03)$ W~m$^{-1}$K$^{-1}$ and $\kappa_{\mathrm{s}} = (144\pm21)$ W~m$^{-1}$K$^{-1}$.

Tong and Majumdar~\cite{Tong2006} describe an alternative formalism to determine the thermal impedance of a multilayer anisotropic substrate. First, the temperature gradient within a single layer is determined by solving a 1D heat conduction equation. Then, to find the temperature at any other measurement depth in the substrate, transfer matrices are used to describe vertical translations and bridging over the interfaces between layers. This formalism is purely one-dimensional but can also be adjusted to account for finite widths of the heat source. For the particular case of a thin film over a semi-infinite bulk layer, the thermal impedance can be determined from the temperature at the surface of the thin film, yielding

\begin{widetext}
\begin{align}
\label{eq:majumdar z_sub}
Z_\mathrm{sub} =
\frac{1}{\kappa_\mathrm{f} \eta_\mathrm{f}}
\frac
{(1+\frac{\kappa_\mathrm{s}\eta_\mathrm{s}}{\kappa_\mathrm{f}\eta_\mathrm{f}} + \frac{\kappa_\mathrm{s}\eta_\mathrm{s}}{h})e^{\eta_\mathrm{f} b_\mathrm{f}}
+ (1-\frac{\kappa_\mathrm{s}\eta_\mathrm{s}}{\kappa_\mathrm{f}\eta_\mathrm{f}} + \frac{\kappa_\mathrm{s}\eta_\mathrm{s}}{h})e^{-\eta_\mathrm{f} b_\mathrm{f}}}
{(1+\frac{\kappa_\mathrm{s}\eta_\mathrm{s}}{\kappa_\mathrm{f}\eta_\mathrm{f}} + \frac{\kappa_\mathrm{s}\eta_\mathrm{s}}{h})e^{\eta_\mathrm{f} b_\mathrm{f}}
- (1-\frac{\kappa_\mathrm{s}\eta_\mathrm{s}}{\kappa_\mathrm{f}\eta_\mathrm{f}} + \frac{\kappa_\mathrm{s}\eta_\mathrm{s}}{h})e^{-\eta_\mathrm{f} b_\mathrm{f}}}
\end{align}
\end{widetext}
where the subscripts $i=\mathrm{f,s}$  denote the film and substrate layers, respectively, and $\eta_i = ({\frac{\kappa_{x_i}}{\kappa_{y_i}}\lambda^2 + \frac{i2\omega}{\alpha_i}})^{1/2}$
is the complex wave vector in the vertical direction. $\lambda$ is the Fourier transform variable in the direction of the finite heater width. In this picture, it is assumed that the heater and the top film are in perfect thermal contact and the heater has negligible thermal mass. However, to account for heat thermal mass and ITR between heater and substrate, one could alternatively consider the heater as one of the layers in the system.

\begin{figure*}
\includegraphics[width=0.99\textwidth]{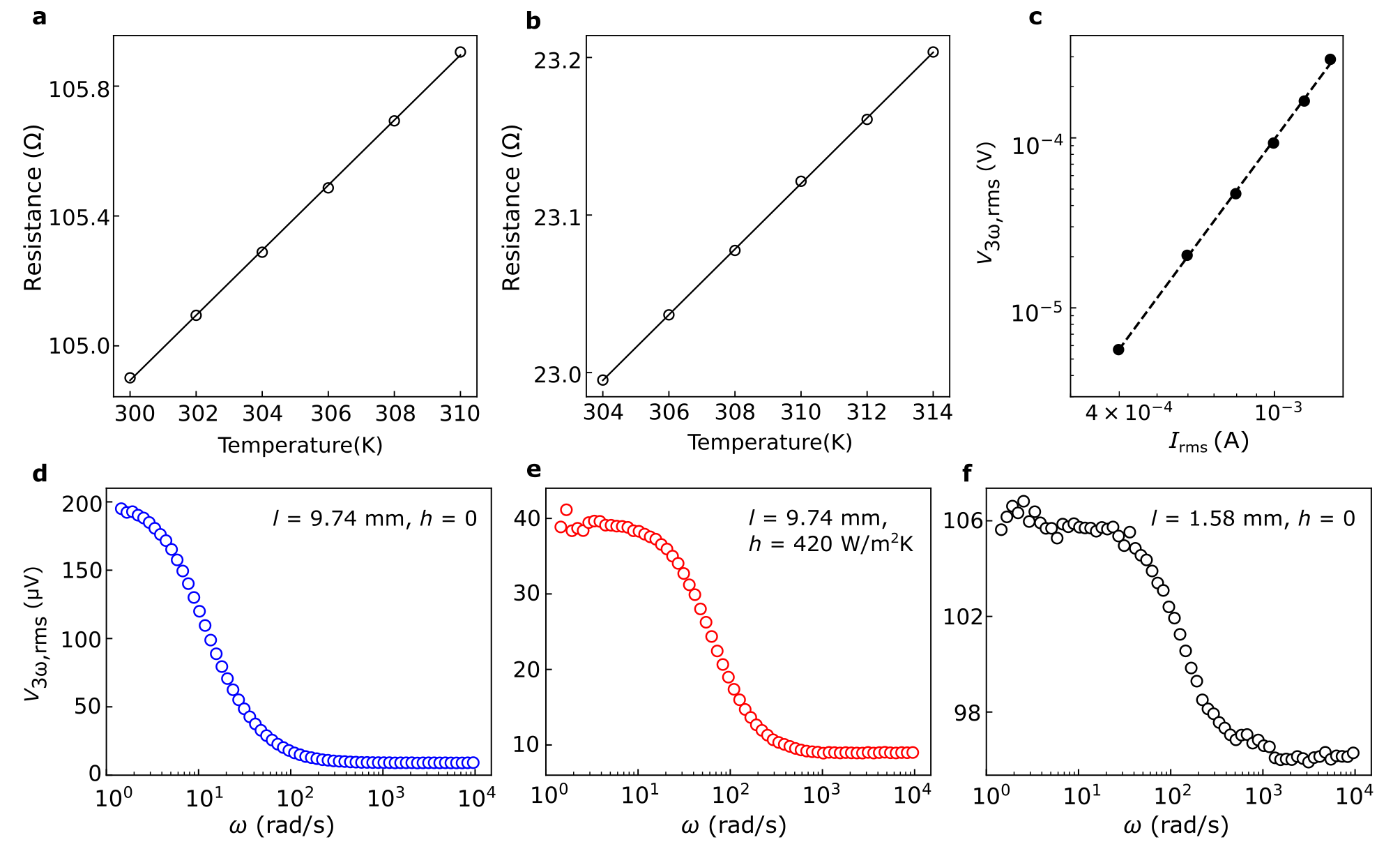}
\caption{\label{fig:S1}3$\omega$ measurement of two carbon nanotube wires. Temperature dependence of electrical resistance for (a) the long wire and (b) the short wire. Linear fit yields $\mathrm{d}R/\mathrm{d}T=(0.100\pm0.003)$~$\Omega$/K for the long wire and $(2.08\pm0.04)\times10^{-2}$~$\Omega$/K for the short wire. (c) $V_{3\omega}$ vs $I_{\mathrm{rms}}$ on log-log scale measured on the long wire under high vacuum. The fitting slope is (3.1$\pm$0.2). (d-f) $3\omega$ voltages measured on the long wire at high-vacuum conditions ($h=0$), the long wire under low-vacuum conditions ($h = 420$~W/(m$^2$K)), and the short wire under high-vacuum conditions, respectively.}
\end{figure*}

\section{3$\omega$ measurement on carbon nanotube wires}
 We conducted 3$\omega$ measurements of two carbon nanotube wire samples (Nanografi) with different lengths. Each wire was suspended between two aluminum nitride (AlN) substrates (Ortech, 0.5~mm thick). Optical microscopy measurements yield diameters of $d_1 = \bar{d_1}\pm\delta d_1=(16.3\pm0.9)$~µm and $d_2 = \bar{d_2}\pm\delta d_2=(17.8\pm 1.7)$~µm for the long and short wires, respectively. These values are mean and standard error determined from 12 measurements taken along the free-standing section of each sample. The long wire sample has a free-standing length of $l_1 = 9.74$~mm, while the short wire sample has a free-standing length of $l_2 = 1.58$~mm. Measurements of the long and short wires were conducted at room temperature conditions (at 300~K and 304~K, respectively). The high thermal conductivity ($>$180~W/(m~K)) of the AlN substrate ensures that the two ends of the free-standing wire sections remain close to the base temperature. The electrical contacts were made through manually placed conductive silver epoxy (EPO-TEK® H20E), which also serves to ensure good thermal contact between the wire and the AlN substrate. An ac current was supplied to the carbon nanotube wire using the internal signal generator of a Zurich Instruments MFLI lock-in amplifier, while the $3\omega$ voltage drop across the suspended part was measured using the two inner contacts.

Fig.~S1 summarizes the measurement results of the two carbon nanotube wire samples. Figs.~S1a and S1b show the temperature dependence of the electrical resistance for the long and short wire, respectively. Fig.~S1c presents a log–log plot of $V_{3\omega,\textrm{rms}}$ versus $I_{\textrm{rms}}$ of the long wire at high vacuum and low frequencies. The plotted $V_{3\omega,\mathrm{rms}}$ values are averaged over the low-frequency range from 0.10 to 2.84~Hz. The extracted slope is (3.1$\pm$0.2), which compares favorably with the expected value of 3. Figs.~S1d-S1f present the $3\omega$ voltage measured on the long wire under high vacuum, the long wire under low vacuum, and the short wire under high vacuum, respectively. To get the data presented in Fig.~3c of the main text, a constant in-phase background due to harmonic distortion of the lock-in output signal has been subtracted from each dataset. We fit the data in Fig.~3c using Eq.~(5) in the main text. For the long wire, the high-vacuum and low-vacuum data are fitted simultaneously, which yields a thermal conductivity of $\kappa = (180\pm20)$~W/(m~K), a volumetric heat capacity of $\rho c_p = (1.30\pm0.15)\times10^6$~J/(m$^3$K) and a heat transfer coefficient for the low-vacuum case of $h = (420\pm30)$ W/(m$^2$K). For the short wire, a best fit yields a thermal conductivity of $\kappa = (140\pm40)$~W/(m~K) and a volumetric heat capacity of $\rho c_p = (2.19\pm0.50)\times10^6$~J/(m$^3$K). For the high-vacuum datasets, the heat transfer coefficient was fixed at $h=0$ during the fitting procedure. The reported values are best fit parameters obtained using the mean diameter measured for each wire ($\bar{d_1}=16.3$~µm for the long wire and $\bar{d_2}=17.8$~µm for the short wire), and the error in each parameter is determined as the maximum difference from the best fit value obtained by refitting the $i=1,2$ (long, short) data with $d_i+\delta d_i$ and $d_i-\delta d_i$. This choice was made as we found the error to be dominated by the significant variation in wire diameter measured along each sample length.

The values of $\kappa$ obtained for the short and long wires
agree with each other, and compare very well with a broad set of reported literature values for carbon nanotube materials.\cite{Marconnet2013,Balandin2011} While the volumetric heat capacity determined from the long wire measurement exceeds the roughly $(0.15-0.35)\times10^6$~J/(m$^3$K) range expected based on the manufacturer's (Nanografi) stated mass density of $300-500$~kg/m$^3$ and room temperature carbon nanotube specific heat values on the order of $500-700$~J/(kg K) reported elsewhere,\cite{Yi1999,Hone2000} it may suggest an actual wire mass density closer to 2000~kg/m$^3$ characteristic of more densely packed nanotubes.\cite{Lekawa2014} However, the volumetric heat capacity obtained for the short wire is unphysical as it exceeds expected room temperature values of both graphite and diamond. Since both wire samples were sourced from the same starting wire, we have no reason to anticipate any significant difference in the intrinsic materials properties among the two samples. Rather, the 3$\omega$ signal from self-heating is significantly weaker relative to the harmonic distortion background of the input signal in the short-wire case compared to the long-wire case (compare Fig. S1f with Figs. S1d–e), suggesting that the constant in-phase subtraction may no longer adequately correct for the harmonic distortion, which is a likely cause of the unphysical heat capacity extracted for the short wire. Ideally, the issue would be avoided altogether by using a signal generator with orders-of-magnitude lower harmonic distortion, and the long-wire results, being less susceptible to this artifact, should therefore be considered more reliable.

\section{Transverse heat transfer model including conductor thermal mass and ITR}

For the schematic geometry shown in Fig.~(1b) of the main text, one often assumes uniform temperature rise along the length of the conductor. With this assumption, the problem reduces to a 2D transverse heat transfer problem from the conductor into the underlying substrate within a cross-sectional plane that has its normal parallel to the axis of the conductor (Fig.~S2a). This heat transfer problem can often be reduced to an effective 1D heat transfer problem. E.g., for the case of a line source conductor on an isotropic, semi-infinite substrate, the heat transfer in the substrate depends only on the radial distance from the conductor.

Based on Cahill,\cite{Cahill1990} the solution for the temperature rise of an infinitely narrow line source on a substrate is
\begin{equation}
\theta_\mathrm{sub} = \frac{Q}{l\pi \kappa_\mathrm{sub}} K_0(qr)
\end{equation}
where $r = \sqrt{x^2 + y^2}$ and $1/q = \sqrt{\alpha_\mathrm{sub}/i\omega}$ with the substrate thermal diffusivity $\alpha_\mathrm{sub} = \kappa_\mathrm{sub}/\rho_\mathrm{sub}c_{p,\mathrm{sub}}$. $K_0$ is the zeroth-order modified Bessel function of the second kind. This narrow line source solution diverges at $r=0$. But since the conductor serves as both the heater and the thermometer, we need to evaluate at $r=0$. Cahill considers the conductor as a spatially uniform heat source with finite width $2b$. One then needs to convolute heat source at every point along the width of the conductor with the solution of the temperature in the sample,
evaluated at $r=0$. Cahill does this by taking the Fourier transform of the above $\theta_\mathrm{sub}$ expression multiplied by the Fourier transform of the uniform heat source (this gives a $\sin(kb)/kb$ term). The final solution after inverse Fourier transform back to real space, and averaging over the conductor width $P=2b$ gives
\begin{equation}
\theta = \theta_\mathrm{sub} = \frac{Q}{l\pi\kappa_\mathrm{sub}} \int_0^\infty \frac{\sin^2(kb)}{(kb)^2 \sqrt{k^2+q^2}} dk = \frac{Q}{l} Z_{\mathrm{sub},l}.
\end{equation}
Here, $Z_{\mathrm{sub},l}$ is the substrate thermal impedance per unit length. It is shown as the red line in Fig.~(S2c).

\begin{figure}
\includegraphics[width=0.45\textwidth]{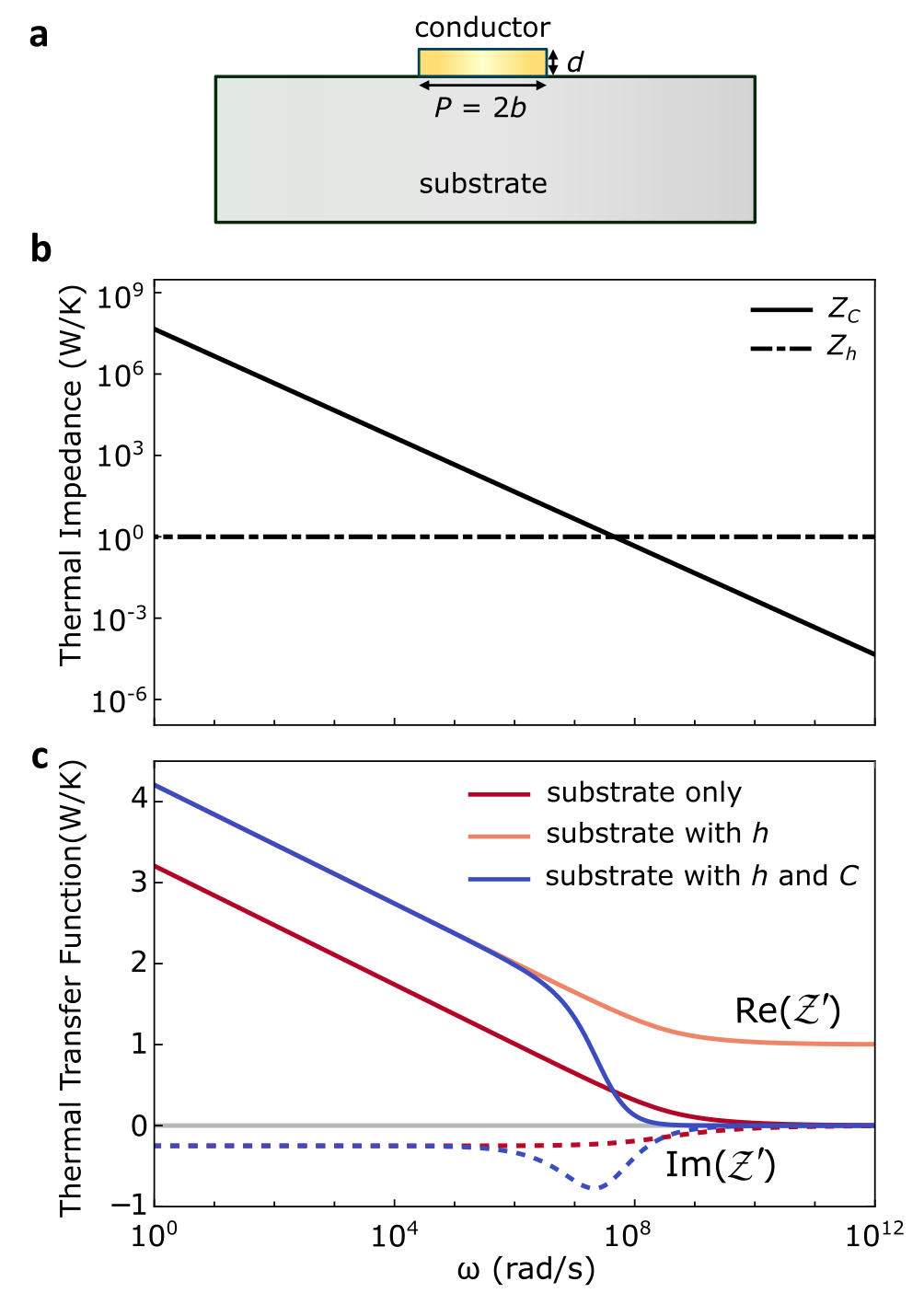}
\caption{\label{fig:S2} (a) Schematic transverse heat transfer geometry. (b) Magnitudes of the thermal impedances entering Eq.~(\ref{eq:zprime}). (c) Thermal transfer function $\mathcal{Z}'$ for three characteristic cases as indicated.}
\end{figure}

We may amend Cahill’s model by an ITR per unit length $1/h_\mathrm{I}P$ and a heater thermal mass per unit length $C_l=\rho c_pS$ (cross-sectional area $S=Pd=2bd$ where $d$ is the thickness of the conductor). Since we have
\begin{equation}
\theta_\mathrm{sub}=\frac{Q'}{l} Z_{\mathrm{sub},l}
\end{equation}
and also
\begin{equation}
\frac{Q'}{l}=h_\mathrm{I}P(\theta-\theta_\mathrm{sub}),
\end{equation}
we can write
\begin{equation}
\frac{Q'}{l}=\frac{h_\mathrm{I}P\theta}{1+h_\mathrm{I}PZ_{\mathrm{sub},l}}.
\end{equation}
Next, consider the heat balance on the heater
\begin{equation}
\frac{Q}{l}=i\omega C_l\theta+\frac{Q'}{l}=i\omega C_l\theta+\frac{h_\mathrm{I}P\theta}{1+h_\mathrm{I}PZ_{\mathrm{sub},l}}.
\end{equation}
Therefore,
\begin{equation}
\label{eq:thetageneral}
\theta=\frac{Q}{l}\frac{1}{i\omega C_l+\left(\frac{1}{h_\mathrm{I}P}+Z_{\mathrm{sub},l} \right)^{-1}}.
\end{equation}
An equivalent expression has been reported by Borca-Tasciuc et al.\cite{borca2001data} Note that this equation corresponds to the thermal mass impedance of the heater per unit length in parallel with the in-series combination of ITR per unit length and $Z_{\mathrm{sub},l}$. Note that in the limiting case where the heater thermal mass is negligible, i.e., $C_l\rightarrow0$, we get
\begin{equation}
\theta=\frac{Q}{l}\left(\frac{1}{h_\mathrm{I}P}+Z_{\mathrm{sub},l} \right).
\end{equation}
This is shown as the orange line in Fig.~(S2c).

From Eq.~(\ref{eq:thetageneral}), if we write $\theta=Q\mathcal{Z}'$, we see that 
\begin{equation}
\label{eq:zprime}
\mathcal{Z'}=\frac{1}{i\omega C+hPl}=\frac{1}{Z_C^{-1}+Z_h^{-1}}.
\end{equation}
is equivalent to the term outside the square brackets in Eq.~(5) of the main text. Eq.~(\ref{eq:zprime}) allows to derive a length-independent criterion for a critical frequency above which the thermal mass of the heater dictates the thermal response:
\begin{equation}
\omega_c'=\frac{|h|Pl}{C}=\frac{|h|P}{\rho c_pS}=\frac{1}{\rho c_pS}\left|\left(\frac{1}{h_\mathrm{I}P}+Z_{\mathrm{sub},l} \right)^{-1}\right|.
\end{equation}
Borca-Tasciuc et al.\cite{borca2001data} have provided an analogous discussion. The transverse heat transfer model discussed in this section does not allow to derive the criterion provided in Eq.~(11) of the main text.

\bibliography{Unified3w}